\documentclass[preprint,3p,sort&compress]{elsarticle}
\usepackage[colorinlistoftodos,textwidth=0.70\marginparwidth]{todonotes}
\usepackage{amsmath}	
\usepackage{amssymb}
\usepackage{color}
\usepackage{graphicx}
\usepackage{slashed}
\makeatletter
\def\ps@pprintTitle{%
 \let\@oddhead\@empty
 \let\@evenhead\@empty
 \def\@oddfoot{\centerline{\thepage}}%
 \let\@evenfoot\@oddfoot}
\makeatother

\def\xiG{\xi_G}
\def\xiW{\xi_W}
\def\xiB{\xi_B}
\def\FeynArts{\texttt{FeynArts}}
\def\DIANA{\texttt{DIANA}}
\def\MATAD{\texttt{MATAD} }
\def\MINCER{\texttt{MINCER} }
\def\BAMBA{\texttt{BAMBA} }
\def\COLOR{\texttt{COLOR} }
\def\lam{\ensuremath{\hat\lambda}}
\def\aS{\ensuremath{a_s}}
\def\aa{\ensuremath{a_1}}
\def\ab{\ensuremath{a_2}}

\DeclareMathOperator{\tr}{tr}

\def\LYukawa{\ensuremath{\mathcal{L}_{\mathrm{Yukawa}}}}

\def\YMf{\ensuremath{Y_{f}}}
\def\YMff{\ensuremath{Y_{ff}}}
\def\YMfff{\ensuremath{Y_{fff}}}

\def\YMu{\ensuremath{Y_{u}}}
\def\YMu{\ensuremath{Y_{u}}}
\def\YMd{\ensuremath{Y_{d}}}
\def\YMl{\ensuremath{Y_{l}}}
\def\YMuu{\ensuremath{Y_{uu}}}
\def\YMdd{\ensuremath{Y_{dd}}}
\def\YMud{\ensuremath{Y_{ud}}}
\def\YMdu{\ensuremath{Y_{du}}}

\def\YMuuu{\ensuremath{Y_{uuu}}}
\def\YMuud{\ensuremath{Y_{uud}}}
\def\YMduu{\ensuremath{Y_{duu}}}
\def\YMdud{\ensuremath{Y_{dud}}}
\def\YMddu{\ensuremath{Y_{ddu}}}
\def\YMudu{\ensuremath{Y_{udu}}}
\def\YMudd{\ensuremath{Y_{udd}}}
\def\YMddd{\ensuremath{Y_{ddd}}}

\def\YMf{\ensuremath{\mathrm{f}}}
\def\YMf{\ensuremath{\mathrm{f}}}
\def\YMff{\ensuremath{\mathrm{ff}}}
\def\YMfff{\ensuremath{\mathrm{fff}}}

\def\YMu{\ensuremath{\mathrm{u}}}
\def\YMu{\ensuremath{\mathrm{u}}}
\def\YMd{\ensuremath{\mathrm{d}}}
\def\YMl{\ensuremath{\mathrm{l}}}
\def\YMuu{\ensuremath{\mathrm{uu}}}
\def\YMdd{\ensuremath{\mathrm{dd}}}
\def\YMud{\ensuremath{\mathrm{ud}}}
\def\YMdu{\ensuremath{\mathrm{du}}}

\def\YMuuu{\ensuremath{\mathrm{uuu}}}
\def\YMuud{\ensuremath{\mathrm{uud}}}
\def\YMduu{\ensuremath{\mathrm{duu}}}
\def\YMdud{\ensuremath{\mathrm{dud}}}
\def\YMddu{\ensuremath{\mathrm{ddu}}}
\def\YMudu{\ensuremath{\mathrm{udu}}}
\def\YMudd{\ensuremath{\mathrm{udd}}}
\def\YMddd{\ensuremath{\mathrm{ddd}}}

\def\YMf{\ensuremath{\mathcal{Y}_{f}}}
\def\YMf{\ensuremath{\mathcal{Y}_{f}}}
\def\YMff{\ensuremath{\mathcal{Y}_{ff}}}
\def\YMfff{\ensuremath{\mathcal{Y}_{fff}}}

\def\YMu{\ensuremath{\mathcal{Y}_{u}}}
\def\YMu{\ensuremath{\mathcal{Y}_{u}}}
\def\YMd{\ensuremath{\mathcal{Y}_{d}}}
\def\YMl{\ensuremath{\mathcal{Y}_{l}}}
\def\YMuu{\ensuremath{\mathcal{Y}_{uu}}}
\def\YMdd{\ensuremath{\mathcal{Y}_{dd}}}
\def\YMud{\ensuremath{\mathcal{Y}_{ud}}}
\def\YMdu{\ensuremath{\mathcal{Y}_{du}}}

\def\YMuuu{\ensuremath{\mathcal{Y}_{uuu}}}
\def\YMuud{\ensuremath{\mathcal{Y}_{uud}}}
\def\YMduu{\ensuremath{\mathcal{Y}_{duu}}}
\def\YMdud{\ensuremath{\mathcal{Y}_{dud}}}
\def\YMddu{\ensuremath{\mathcal{Y}_{ddu}}}
\def\YMudu{\ensuremath{\mathcal{Y}_{udu}}}
\def\YMudd{\ensuremath{\mathcal{Y}_{udd}}}
\def\YMddd{\ensuremath{\mathcal{Y}_{ddd}}}
\def\Yu{\ensuremath{\tr [\YMu]}}
\def\Yd{\ensuremath{\tr [\YMd]}}
\def\Yl{\ensuremath{\tr [\YMl]}}
\def\Yuu{\ensuremath{\tr [\YMuu]}}
\def\Ydd{\ensuremath{\tr [\YMdd]}}
\def\Yud{\ensuremath{\tr [\YMud]}}

\def\Yuuu{\ensuremath{\tr [\YMuuu]}}
\def\Yuud{\ensuremath{\tr [\YMuud]}}
\def\Yudd{\ensuremath{\tr [\YMudd]}}
\def\Yddd{\ensuremath{\tr [\YMddd]}}

\newcommand{\MS}{{\ensuremath{\overline{\mathrm{MS}}}}}
\def\z#1{{\zeta_{#1}}}
\def\LYukawa{\ensuremath{\mathcal{L}_{\mathrm{Yukawa}}}}

\newcommand{\NGen}{\ensuremath {n_G}}

\def \eps {\epsilon}

\journal{Physics Letters B}
\begin{document}
\begin{frontmatter}
\begin{flushright}\footnotesize
\texttt{HU-Mathematik-2014-17}\\
\texttt{HU-EP-14/27}\\
\vspace{0.5cm}
\end{flushright}

\title{Three-loop SM beta-functions for matrix Yukawa couplings}
\author[a]{A.~V.~Bednyakov}
\ead{bednya@theor.jinr.ru}
\author[a]{A.~F.~Pikelner}
\ead{andrey.pikelner@cern.ch}
\author[b,c]{V.~N.~Velizhanin}
\ead{velizh@thd.pnpi.spb.ru}

\address[a]{Joint Institute for Nuclear Research,\\
 141980 Dubna, Russia}

\address[b]{Institut f{\"u}r  Mathematik und Institut f{\"u}r Physik, Humboldt-Universit{\"a}t zu Berlin, \\
IRIS Adlershof, Zum Gro\ss en Windkanal 6, 12489 Berlin, Germany}

\address[c]{Theoretical Physics Division, Petersburg Nuclear Physics Institute, \\
Orlova Roscha, Gatchina, 188300 St.~Petersburg, Russia}

\begin{abstract}
We present the extension of our previous results for three-loop Yukawa coupling beta-functions to the case
of complex Yukawa matrices describing the flavour structure of the SM. 
The calculation is carried out in the context of unbroken phase of the SM with the help of the \texttt{MINCER} program
in a general linear gauge and cross-checked by means of \texttt{MATAD}/\texttt{BAMBA} codes.
In addition, ambiguities in Yukawa matrix beta-functions are studied.

\end{abstract}

\end{frontmatter}

It is an important property of the SM that all the particle masses are related to the corresponding couplings of the Higgs boson, 
	discovered recently in the LHC experiments~\cite{Aad:2012tfa,Chatrchyan:2012ufa}.  
	Careful experimental investigation of the Higgs decay modes shows consistency with the prediction of the SM~\cite{ATLAS:2013oma,CMS:yva}.
	This kind of studies is complemented by the experiments aimed to shed light on the SM flavour structure (see Ref.~\cite{Gershon:2013aca}
	and references therein).   
In spite of the fact that the SM provides a consistent description of the processes involving transitions between
	different fermion generations, the origin of flavour physics is still unclear. 
Most of the observable flavour effects in the SM are encoded in the Cabibbo-Kobayashi-Maskawa (CKM) matrix which enters into the tree-level
charged quark currents.\footnote{In this paper, we do not consider mixing of massive neutrinos described by Pontecorvo-Maki-Nakagawa-Sakata (PMNS) matrix (see, e.g., Ref.~\cite{Bilenky:2012qb} for review).}
The CKM matrix originates from matrix Yukawa couplings after transition from weak (interaction) eigenstates to the mass basis. 
It is obvious that general complex matrices involve a lot of unphysical parameters which can be ``rotated'' away by unitary transformations.
However, it is sometimes convenient to study this general structure in view of possible New Physics which can potentially explain the observed
	hierarchy in quark masses and mixing (see, e.g., Ref.~\cite{Fritzsch:1999ee}).

	This article concludes the series of our papers on three-loop beta-functions in the SM with complex Yukawa matrices~\cite{Bednyakov:2012rb,Bednyakov:2013cpa}.  
One-
and two-loop results for SM beta-functions have been known for quite a long time~\cite{Gross:1973id,Politzer:1973fx,Jones:1974mm,Tarasov:1976ef,Caswell:1974gg,Egorian:1978zx,Jones:1981we,Fischler:1981is,Machacek:1983tz,Machacek:1983fi,Luo:2002ti,Jack:1984vj,Gorishnii:1987ik,Arason:1991ic}
and are summarized in Ref.~\cite{Luo:2002ey}, in which the renormalization group equations (RGE) are given in the matrix form. 
The three-loop gauge-coupling beta-functions with the full flavour structure were calculated for the first time 
in Ref.~\cite{Mihaila:2012pz} and confirmed later by our group \cite{Bednyakov:2012rb}. 
The beta-functions for the parameters of the Higgs potential in the case of complex Yukawa matrices were considered in Ref.~\cite{Bednyakov:2013cpa}
(for a flavour diagonal case, see Refs.~\cite{Chetyrkin:2013wya,Bednyakov:2013eba}). 
It is interesting to note that the results for three-loop RGE in the Minimal Supersymmetric Standard Model were found
by means of the supergraph formalism \cite{Grisaru:1979wc} about ten years ago~\cite{Jack:2004ch}.

In order to obtain the results for matrix beta-functions, we use \FeynArts~\cite{Hahn:2000kx} and \DIANA~\cite{Tentyukov:1999is}. 
In both cases, we extend the corresponding model files to account for explicit flavour indices and develop simple
	routines for dealing with them.
Since the matrix couplings do not pose any additional problems in $\gamma_5$ treatment, we will not discuss 
it here and only refer to Refs. \cite{Chetyrkin:2012rz,Bednyakov:2012en}.

As in our previous studies, the calculation is carried out in an almost automatic way with the help of different infrared rearrangement (IRR)~\cite{Vladimirov:1979zm}
	procedures implemented in our codes.
The utilized IRR prescriptions consist in either setting all but two external momenta to zero or introducing an auxiliary mass parameter $M$ in every propagator
	and the subsequent expansion in all external momenta~\cite{Misiak:1994zw,Chetyrkin:1997fm}. 
In the first case, one uses \MINCER~\cite{Gorishnii:1989gt,Larin:1991fz} to compute massless propagator-type integral.\footnote{No spurious IR divergences are generated in Yukawa vertices if one neglects
	the momentum entering into the scalar leg.} 
In the second approach, the fully massive vacuum integrals are evaluated by means of the \MATAD~package~\cite{Steinhauser:2000ry} 
	or \BAMBA~code developed by V.N.~Velizhanin.
For the color algebra the FORM package \COLOR~\cite{vanRitbergen:1998pn} is utilized. 
	
Let us briefly specify our notation and renormalization procedure.
The full Lagrangian of the ``unbroken'' (= massless) SM which was used in this calculation is given
	in our previous paper~\cite{Bednyakov:2012rb}.
For the reader's convenience we present here only the terms describing the fermion-Higgs interactions
	\begin{eqnarray}
	  \LYukawa &=& - \bigg(
	Y^{ij}_u (Q^L_i \Phi^c) u^R_j
	+ Y^{ij}_d (Q^L_i \Phi) d^R_j  + Y^{ij}_l (L^L_i \Phi) l^R_j
	+ \mathrm{h.c.} \bigg).
	  \label{eq:yukawa_lag} 
  \end{eqnarray}
Here $Y_{u,d,l}$ are the Yukawa matrices.
The Higgs doublet $\Phi$ and its charge-conjugated counterpart $\Phi^c=i\sigma^2\Phi^\dagger$ have 
  hypercharges $Y_W = 1$ and $Y_W=-1$, respectively.\footnote{The hypercharge normalization is fixed by its relation to electric charge $Q = T_3 + Y_W/2$ and weak isospin $T_3$.}
The left-handed quark and lepton $SU(2)$ doublets, $Q^L_i$ and $L^L_i$, carry flavour indices $i=1,2,3$.  
The same is true for the $SU(2)$ singlets corresponding to the right-handed SM fermions  
	$u_i^R$, $d_i^R$ and $l^R_i$.
It is worth mentioning that the matrix element $Y^{ij}_u$ describes the transition 
  of the right-handed up-type quark of the $j$-th generation to the left-handed quark (either up- or down-type) 
  of the generation $i$. 
Conversely, the matrix element $Y^{\dagger,ij}_u$ corresponds to the transition of the left-handed quark from the
  doublet $Q_j$ to the right-handed up-type quark $u_i$. 

In the Lagrangian \eqref{eq:yukawa_lag} we assume that the bare fermion fields and Yukawa couplings are expressed in terms of 
the corresponding running quantities defined in the \MS~renormalization scheme by means of 
($f=u,d,l$)
\begin{eqnarray}
	\left(Q^L\right)_{\mathrm{Bare}}  =  \left[Z^{1/2}_{Q}\right] Q^L \mu^{-\eps}, \quad  
	&\left(u^R\right)_{\mathrm{Bare}}  =  \left[Z^{1/2}_{u}\right] u^R \mu^{-\eps},& 
	\quad \left(d^R\right)_{\mathrm{Bare}}  =  \left[Z^{1/2}_{d}\right] d^R \mu^{-\eps}, 
	\label{eq:quarks_ren}\\
	\left(L^L\right)_{\mathrm{Bare}}  =  \left[Z^{1/2}_{L}\right] L^L \mu^{-\eps}, \quad 
	& \left(l^R\right)_{\mathrm{Bare}}  =  \left[Z^{1/2}_{l}\right] l^R \mu^{-\eps}, &
	\quad \left(Y_f\right)_{\mathrm{Bare}}  =  \left(Y_f + \Delta Y_f \right) \mu^{\epsilon},  
	\label{eq:leptons_ren}
\end{eqnarray}
where the generation indices are  suppressed.
The renormalization constants $Z=(Z_Q^{1/2}, Z_L^{1/2}, Z_u^{1/2}, Z_d^{1/2}, Z_l^{1/2})$ and $\Delta Y_f$ 
	are $3\times 3$ matrices in the flavour space and  can be decomposed as 
\begin{eqnarray}
	Z &=& 1 + \sum_{l=1}^\infty \delta Z^{(l)}, \qquad \delta Z^{(l)} = \sum_{n=1}^l \frac{c^{(l,n)}}{\epsilon^n},
	\label{eq:Z_decomposition}\\
	\Delta Y_f &=& \sum_{l=1}^\infty \left[\Delta Y_f^{(l)}\right], 
	\qquad \left[\Delta Y_f^{(l)}\right] = \sum_{k=1}^l \frac{1}{\epsilon^k} \left[\Delta Y_f^{(l,k)} \right] 
	\label{eq:dY_decomposition}
\end{eqnarray}
	with $\delta Z^{(l)}$ and $\Delta Y_f^{(l)}$ being $l$ - loop contributions 
	and $\epsilon = (D-4)/2$ corresponding to the parameter of dimensional
	regularization.

As it was mentioned above, we used two approaches to IRR. 
Let us consider the first one, when all Feynman integrals are converted to massless propagators and it is convenient
to use multiplicative renormalization of Green functions. 
Contrary to the second case, when a new auxiliary mass is introduced, no new parameter appears in the problem, 
so that the matrix renormalization constants can be recursively obtained via the relations
	\begin{eqnarray}
		\Gamma^{(2)}_{f,\mathrm{Ren}}\left(\frac{k^2}{\mu^2}, a_{i}, Y_f\right) 
		= \left[Z_f^{1/2}\right]^\dagger
		\Gamma^{(2)}_{f,\mathrm{Bare}}\left(k^2, a_{i,\mathrm{Bare}}, 
		Y_{f,\mathrm{Bare}}, \epsilon \right) \left[Z_f^{1/2}\right]
	\label{eq:ren_2pt} \\
	\Gamma^{(3)}_{\bar{f}'f\phi,\mathrm{Ren}}\left(\frac{k_i^2}{\mu^2}, a_{i}, Y_f\right) 
		= \left[Z_{f'}^{1/2}\right]^\dagger 
		\Gamma^{(3)}_{\bar{f'}f\phi,\mathrm{Bare}}\left(k_i^2, a_{i,\mathrm{Bare}}, 
		Y_{f,\mathrm{Bare}}, \epsilon \right) \left[Z_f^{1/2}\right] Z_\phi^{1/2}.
	\label{eq:ren_3pt} 
	\end{eqnarray}
In these equations $\Gamma_f^{(2)}$ corresponds
	to the one-particle irreducible (1PI)  two-point Green functions
	for the fermion $f$ (left-handed or right-handed).
	The three-point 1PI vertex $\Gamma^{(3)}_{\bar {f'}f\phi}$ describes the transition of the right-handed fermion $f$
	to the left-handed fermion $f'$ due to interactions with the Higgs boson $\phi$, 
	either neutral $\phi=h,\chi$ or charged $\phi=\phi^\pm$. 
The Green functions are normalized in such a way, that at the tree-level 
$\Gamma_f^{(2)} = 1$ and $\Gamma^{(3)}_{\bar{f'}f\phi} = Y_f$. 
	As in our previous papers \cite{Bednyakov:2012rb,Bednyakov:2012en}, we define the following quantities:
\begin{equation}
	 a_i   =   \left(\frac{5}{3} \frac{g_1^2}{16\pi^2}, \frac{g_2^2}{16\pi^2}, \frac{g_s^2}{16\pi^2},
	 \frac{\lambda}{16\pi^2}, \xiG, \xiW, \xiB \right)
	\label{eq:coupl_notations}
\end{equation}
	with $g_s$, $g_2$, $g_1$ being $SU(3)\times SU(2) \times U(1)$ gauge couplings,
	$\lambda$ --- Higgs self-coupling and $\xiG$, $\xiW$, $\xiB$ corresponding to the gauge-fixing parameters
	in the unbroken SM.
In addition, the following abbreviations ($f = u, d, l$)
\begin{eqnarray}
	\YMf \equiv  \frac{ Y_f Y_f^\dagger}{16\pi^2},
	 \qquad 
	\YMff  &\equiv &  \frac{ Y_f Y_f^\dagger Y_f Y_f^\dagger}{(16\pi^2)^2}, 
	\qquad
	\YMfff \equiv  \frac{ Y_f Y_f^\dagger Y_f Y_f^\dagger Y_f Y_f^\dagger}{(16\pi^2)^3},
\nonumber\\
	  \YMud =  \frac{Y_u Y_u^\dagger Y_d Y_d^\dagger}{(16\pi^2)^2},
	  \qquad 
	  \YMdu & \equiv &  \frac{Y_d Y_d^\dagger Y_u Y_u^\dagger}{(16\pi^2)^2},
	 \qquad 
	 \nonumber\\
	  \YMuud \equiv  \frac{ Y_u Y_u^\dagger Y_u Y_u^\dagger Y_d Y_d^\dagger}{(16\pi^2)^3}, 
	  \qquad
	  \YMudu & \equiv &  \frac{ Y_u Y_u^\dagger Y_d Y_d^\dagger Y_u Y_u^\dagger}{(16\pi^2)^3},
	  \qquad 
	  \YMduu  \equiv   \frac{ Y_d Y_d^\dagger Y_u Y_u^\dagger Y_u Y_u^\dagger}{(16\pi^2)^3},\nonumber\\
	  \YMudd \equiv  \frac{ Y_u Y_u^\dagger Y_d Y_d^\dagger Y_d Y_d^\dagger}{(16\pi^2)^3},
	  \qquad
	  \YMddu & \equiv &  \frac{ Y_d Y_d^\dagger Y_d Y_d^\dagger Y_u Y_u^\dagger}{(16\pi^2)^3},
	   \qquad 
	  \YMdud  \equiv   \frac{ Y_d Y_d^\dagger Y_u Y_u^\dagger Y_d Y_d^\dagger}{(16\pi^2)^3}
	  \label{eq:yukawa_not}
\end{eqnarray}
	will be used for the Yukawa matrix products.

It is worth pointing out that 
	in the absence of Yukawa interactions \eqref{eq:yukawa_lag} the SM Lagrangian is invariant under 
	an accidental global flavour symmetry\footnote{Corresponding to $SU(2)$-compatible rotations of the left-handed ($Q$ and $L$) and right-handed ($u$, $d$, and $l$) fermion fields.}
	$U(3)_Q\times U(3)_u \times U(3)_d \times U(3)_L \times U(3)_l$. 
Due to this, we have an equivalence relation
\begin{eqnarray}
	&&\left( Y_u, Y_d, Y_l \right) 
 \Leftrightarrow 
	\left( Y'_u, Y'_d, Y'_l \right) = 
	\left( V_Q Y_u V_u^\dagger, 
	       V_Q Y_d V_d^\dagger,  
	       V_L Y_l V_l^\dagger  
	\right),
	\qquad
	V_f \in U(3)_f, \quad f = Q,L,u,d,l, 
	\label{eq:equiv_classes}
\end{eqnarray}
implying that the Lagrangians~\eqref{eq:yukawa_lag} with couplings $\left( Y_u, Y_d, Y_l \right)$ and $\left( Y'_u, Y'_d, Y'_l \right)$ 
	lead to the same physics \cite{Santamaria:1993ah},
since one can always compensate the factors $V_f$ by appropriate basis change $f \to f' = V_f f$ without affecting the rest of the SM Lagrangian.
It is easy to notice that Eq.~\eqref{eq:equiv_classes} entails  equivalence relations for the matrix products \eqref{eq:yukawa_not}, i.e., 
	$\mathcal{Y}_{q\dots q'} \Leftrightarrow \mathcal{Y}'_{q \dots q'} = V_Q \mathcal{Y}_{q \dots q'} V_Q^\dagger$.
This property of the Lagrangian will be important in our discussion of the ambiguities in the Yukawa matrix beta-functions.

The left-hand side (LHS) of Eqs.\eqref{eq:ren_2pt}-\eqref{eq:ren_3pt} is finite when the parameter $\epsilon$ tends
	to zero.
This allows us to find the expressions for certain combinations of \MS-renormalization constants by canceling recursively the poles 
	in $\epsilon$, which appear in the right-hand side (RHS) of the same equations.
From the fermion self-energies $\Gamma_f^{(2)}$ we extract a hermitian combination 
	$Z_{2,f} = \left[Z^{1/2}_f\right]^\dagger Z^{1/2}_f$.
	Taking the square root operation in perturbation theory one fixes the hermitian part (denoted by $\tilde{Z}_f^{1/2}$) of $Z^{1/2}_f$: 
\begin{eqnarray}
	\tilde{Z}_f^{1/2} & = & 1 + \frac{1}{2} \delta Z^{(1)}_{2,f} 
	+ \frac{1}{2} \left( \delta Z^{(2)}_{2,f} - \frac{1}{4} \delta Z^{(1)}_{2,f} \delta Z^{(1)}_{2,f} \right) \nonumber\\
	& & \phantom{1} + \frac{1}{2}\left( \delta Z^{(3)}_{2,f} 
			- \frac{1}{4} \left( \delta Z^{(1)}_{2,f} \delta Z^{(2)}_{2,f} 
					   + \delta Z^{(2)}_{2,f} \delta Z^{(1)}_{2,f} \right)
			+ \frac{1}{8} \delta Z^{(1)}_{2,f} \delta Z^{(1)}_{2,f} \delta Z^{(1)}_{2,f}\right) + \dots, 
	\label{eq:ren_ferm_herm}
\end{eqnarray}
	where only the terms relevant for our three-loop calculations are retained. 
It turns out that at three loops  the hermitian factors $\tilde{Z}^{1/2}_f$ used in place of $Z^{1/2}_f$ 
	give rise to infinite expressions (in the limit $\epsilon \to 0$) 
	for the matrix anomalous dimension $\gamma_f$ of renormalized quark fields $\mathfrak{F}_f=Q^L,u^R,d^R$  defined as
\begin{eqnarray}
	\gamma_{f} \cdot \mathfrak{F}_f(\mu) \equiv \left.\frac{d }{d \ln\mu^2} \mathfrak{F}_f(\mu,\eps)\right|_{\eps=0} 
	= - \left(Z_f^{-1/2} \frac{d}{d\ln\mu^2} Z_f^{1/2} \right) \cdot \mathfrak{F}_f.
	\label{eq:fermion_anom_dim_def}
\end{eqnarray}
In other words, pure hermitian renormalization constants do not satisfy 
	the pole equations \cite{'tHooft:1972fi} for the $u$- and $d$-quark. 
To circumvent this problem, we introduced the following unitary factors:
\begin{eqnarray}
	\bar{Z}^{1/2}_{Q_L} & = & 
		1  -  
		 \frac{a_1 h^3}{320} \left(\frac{1}{6 \epsilon^2} - \frac{1}{\epsilon^3} 
		 \right) 
		 \bigg[\YMu,\YMd\bigg]
		+ \frac{h^3}{64} \left(\frac{1}{6 \epsilon^2} + \frac{1}{\epsilon^3} \right) 
		\bigg\{ \YMu - \YMd, \bigg[ \YMu, \YMd \bigg] \bigg\}
		,	 
	\label{eq:ren_ferm_nonherm_L} \\
	\bar{Z}^{1/2}_{u_R} & = & 
	1  - \frac{h^3}{32}  \left(\frac{1}{6 \epsilon^2} - \frac{1}{\epsilon^3} \right) 
		Y^\dagger_u \bigg[
		\YMu, \YMd 
		\bigg] Y_u 	
		,	
		\quad
	\bar{Z}^{1/2}_{d_R}  =  
	1  + \frac{h^3}{32}  \left(\frac{1}{6 \epsilon^2} - \frac{1}{\epsilon^3} \right) 
		Y^\dagger_d \bigg[
			\YMu, \YMd
		\bigg] Y_d,	
	\label{eq:ren_ferm_nonherm_q_R}
\end{eqnarray}
	where the commutator $\left[\YMu,\YMd\right]$ is an 
	anti-hermitian matrix, which is a measure of whether $Y_u$ and $Y_d$ can be diagonalized simultaneously,\footnote{It is interesting to note that  
	$\det \left[ \YMu, \YMd \right] = - 2 T(y_u) B(y_d) J$, where 
	$T(y_u) = (y_t^2 - y_u^2) (y_t^2 - y_c^2) (y_c^2 - y_u^2)$, 
	$B(y_d) = (y_b^2 - y_d^2) (y_b^2 - y_s^2) (y_s^2 - y_d^2)$, 
	with $y_f$ and $J$ being the Yukawa coupling for the quark mass eigenstate $f$ and the Jarlskog invariant \cite{Jarlskog:1985cw}, respectively.
	The latter can be expressed in terms of CKM matrix elements $V_{ij}$, e.g., $J=\mathrm{Im}(V_{11} V_{22} V^*_{12} V^*_{21})$, and measures
	CP-violation in the SM.} 
	and 
	$h^l$ is used to indicate $l$-loop contribution. 
	Due to the unbroken $SU(2)$ invariance, we have $\bar{Z}^{1/2}_{u_L}=\bar{Z}^{1/2}_{d_L} = \bar{Z}^{1/2}_{Q_L}$.
	The factors $\tilde{Z}^{1/2}$ and $\bar{Z}^{1/2}$ combine to form $Z^{1/2}$ 
\begin{equation}
	Z_f^{1/2}  = \bar{Z}_f^{1/2} \tilde{Z}_f^{1/2}, \qquad 
	\left[Z_f^{1/2}\right]^\dagger   = \tilde{Z}_f^{1/2} \left[\bar{Z}_f^{1/2}\right]^\dagger, 
	\qquad \left[\bar{Z}_f^{1/2}\right]^\dagger \bar{Z}_f^{1/2} = 1, 
	\qquad Z_{2,f} = \left[\tilde{Z}_f^{1/2}\right]^2.
	\label{eq:ren_ferm_all}
\end{equation}
The renormalization constants for other SM quantum fields, e.g., $Z^{1/2}_\phi$ required to define
	\eqref{eq:ren_3pt}, can be easily obtained from the corresponding two-point functions.
In order to find the renormalization constant for the Yukawa matrix $Y_f$, we use \eqref{eq:ren_3pt}. 
After two-loop renormalization of RHS we are left with the divergence, which should be canceled
	by the three-loop part of the vertex counter-term
\begin{equation}
	Z_{\bar{f}'f\phi} Y_f \equiv \left[Z_{f'}^{1/2}\right]^\dagger \left(Y_f + \Delta Y_f\right)
		\left[Z_f^{1/2}\right] Z_\phi^{1/2},
	\label{eq:yuk_vertex_ct}
\end{equation}
	originating from the Lagrangian \eqref{eq:yukawa_lag}. 
It is worth noticing that it is always possible to factorize $Y_f$ ($Y_f^\dagger$) 
	from the right (left) of the considered matrix three-point Green functions involving 
	incoming (outgoing) right-handed fermion $f$. 
From Eq.~\eqref{eq:yuk_vertex_ct} one can deduce that
\begin{equation}
	Y_f + \Delta Y_f =   
	\left[Z_{f'}^{-1/2}\right]^\dagger Z_{\bar{f}'f\phi} Y_f \left[Z_f^{-1/2}\right] Z_\phi^{-1/2}
	 = 
	 \bar{Z}_{f'}^{1/2} \left[\tilde{Z}_{f'}^{-1/2} Z_{\bar{f}'f\phi} Y_f \tilde{Z}_f^{-1/2}Z_\phi^{-1/2}\right] \left[\bar{Z}_f^{1/2}\right]^\dagger
	\label{eq:yuk_ren}
\end{equation}
In our calculation we used  $\Gamma^{(3)}_{\bar{f} f \phi}$ with $f=u,d,l$ and $\phi=h,\chi$ 
	for extraction of $\Delta Y_f$. 
Moreover, the vertex $\Gamma^{(3)}_{\bar{u} d \phi^+}$ was also considered for additional 
	verification of the correctness of the results. 
The Green function with charged would-be goldstone $\phi^+$ allows us to find the counter-terms for both
	$Y^\dagger_u$ ($\bar{u}_R d_L \phi^+$) and $Y_d$ ($\bar{u}_L d_R \phi^+$).  
Irrespectively of the considered vertex the results for $Y_f$ turn out to be the same.

The matrix Yukawa beta-functions $\beta_{Y_f}$ are defined by
\begin{eqnarray}
	\beta_{Y_f} Y_f \equiv \frac{d Y_f(\mu,\epsilon)}{d \ln \mu^2}\bigg|_{\epsilon=0} 
	\label{eq:Yuk_beta_def}
\end{eqnarray}
	and can be found from the relation between bare and renormalized couplings by differentiation
\begin{eqnarray}
	0 = \frac{d}{d \ln \mu^2} \left(Y_f\right)_{\mathrm{Bare}}  & = &  
    \left(-\frac{\epsilon}{2} Y_f + \beta_{Y_f} Y_f + \frac{d}{d \ln \mu^2} (\Delta Y_f) \right) \mu^{\epsilon} + \frac{\epsilon}{2} \left( Y_f + \Delta Y_f\right) \mu^\epsilon,
	 \label{eq:Yuk_beta_deriv}
\end{eqnarray}
	from which one deduces 
\begin{eqnarray}
	\beta_{Y_f} Y_f = \sum_{l=1}^{\infty} \left[ 
	a_i \frac{\partial}{\partial a_i}  + \frac{1}{2} 
	 \sum_{f'=u,d,l}\left(Y^{ij}_{f'}\frac{\partial}{\partial Y_{f'}^{ij}}  + 
	Y^{\dagger,ij}_{f'} \frac{\partial}{\partial Y_{f'}^{\dagger,ij}} \right)	
	- \frac{1}{2}
	\right] \Delta Y_f^{(l,1)}
	= 
	\sum_{l=1}^{\infty} l \cdot \Delta Y_f^{(l,1)}, 
	\label{eq:Yuk_beta_calc}
\end{eqnarray}
	assuming the validity of the corresponding pole equations.

It is interesting to mention that the expressions for the matrix renormalization constants discussed in this article were obtained 
	for the first time by means of the described procedure. 
The computer setup 
	(\FeynArts \cite{Hahn:2000kx} for diagram generation +  \MINCER for integral computations) 
	was the same that we used in our first two papers \cite{Bednyakov:2012rb,Bednyakov:2012en} 
	on the three-loop SM beta-function.
The obtained expressions for $\Delta Y_f$ were free from gauge-parameter dependence. 
However, in this calculation we did not take into account the unitary factors $\bar{Z}^{1/2}_f$ and were unable 
	to satisfy the pole equations for the quark fields and Yukawa couplings.

In order to figure out the problem and to cross-check the obtained results, we made use of another established setup 
	(\DIANA + \MATAD / \BAMBA), which is based on the second mentioned approach to IRR.
The crucial difference in the renormalization procedure is that one calculates renormalization constants 
for  Green function $\Gamma$ via
\begin{equation}
	Z_\Gamma = 1 - \mathcal{K} \mathcal{R}' \Gamma,
	\label{eq:KR}
\end{equation}
with $\mathcal{R}'$ being the incomplete $\mathcal{R}$-operation without the last subtraction (see, e.g., Ref.~\cite{Kazakov:2008tr}), 
	and $\mathcal{K}$ extracts the singular part in $\epsilon$. 
The implementation of the method requires introduction of explicit
	counter-term insertions corresponding to all SM fields and parameters and, in addition, 
	to the auxiliary mass $M$.
By means of this procedure the same result for renormalization constants of the considered
	two- and three-point Green functions was obtained. 
	This ensures the correctness of the corresponding expressions.

After a careful study of the employed procedures we have found that the square root operation, 
	which was used to find $Z_f^{1/2}$, has an ambiguity.
The latter was utilized and the unitary factors $\bar{Z}_f^{1/2}$ were introduced.
It is worth mentioning that these factors themselves do not satisfy pole equations so that
	one can not define the finite anomalous dimensions 
	$\bar \gamma_f = - \bar Z_f^{-1/2} \dot{\bar Z}_f^{1/2}$.

A comment on the remaining ambiguity is in order since one can introduce additional
	unitary factors involving the first poles in $\epsilon$. 
For example, 
	we can  multiply the obtained renormalization constant
	$Z_{Q_L}^{1/2}$  by a factor ($A_1$ is an arbitrary constant) 
\begin{eqnarray}
	\mathcal{Z}_{Q_L} = 1 & + & \frac{h^2}{\epsilon} A_1 \bigg[ 
		  \YMu,\YMd
		\bigg] 			
		\nonumber\\
		& + & 
		 \frac{h^3}{\epsilon^2}
		A_1
		\left[
		\frac{
		2 \alpha^u_u  + \alpha^d_u
		}{3}
		\bigg\{
		\YMu,
		\bigg[ 
		\YMu, \YMd
		\bigg] 
		\bigg\}
		+
		\frac{
		2 \alpha^d_d  + \alpha^u_d
	}{3}
		  \bigg\{
		  \YMd,
		  \bigg[
		\YMu, \YMd
		  \bigg]
		  \bigg\}
		+
		\frac{
			2 \alpha_0^d  
		+	2 \alpha_0^u
	}{3}
		  \bigg[
		\YMu, \YMd
		  \bigg]
		  \right]
	\label{eq:zql_ambiguity}
\end{eqnarray}
	without any effect on the propagator ($Z^{1/2\dagger}_{Q_L} Z^{1/2}_{Q_L} - 1$) and vertex $(Z_{\bar{f}'f\phi} - 1) Y_f$ counter-terms. 
The coefficient of $h^3/\epsilon^2$ in \eqref{eq:zql_ambiguity} is determined 
	from pole equations and ensures 
	the finiteness of the 
	corresponding anomalous dimension 
	$\gamma_L' \equiv - \mathcal{Z}_{Q_L}^\dagger \dot{\mathcal{Z}}_{Q_L} = 2 A_1 h^2 ( \YMud - \YMdu )$
	up to three loops.
The coefficients $\alpha_i^u$ and $\alpha_i^d$ enter into one-loop beta-functions for $Y_u$ and $Y_d$:
\begin{eqnarray}
	\beta_{Y_f} & = &  \alpha^f_u \cdot \YMu + \alpha^f_d \cdot \YMd + \alpha^f_0,\qquad f = u,d,
	\label{eq:beta_yuk_1l}
\end{eqnarray}
	and in the SM 
\begin{eqnarray}
	  \alpha^u_u & = & \alpha^d_d = - \alpha^u_d = - \alpha^d_u = \frac{3}{4}, \quad
	  \begin{pmatrix} 
		  	         \alpha^u_0  \\
		  		 \alpha^d_0  	
	\end{pmatrix}
		= 
				- 4 \aS - \frac{9}{8} \ab  + \frac{ 3 \Yu + 3 \Yd + \Yl }{2}
	- \begin{pmatrix}
		\frac{17}{40} \\
		\frac{1}{8}
	  \end{pmatrix}
	  \aa
	\label{eq:beta1l_yuk_coeff}
\end{eqnarray}
It is clear that the substitution $Z_{Q_L}^{1/2} \to \mathcal{Z}_{Q_L} Z_{Q_L}^{1/2}$ in \eqref{eq:quarks_ren} will modify the anomalous dimension for the left-handed quarks and the Yukawa coupling beta-functions
	in the following way\footnote{In the general case we have $\beta_{Y_f} Y_f  \to \beta_{Y_f} Y_f + \gamma_L' Y_f - Y_f \gamma_f'$ 
	with $\gamma_f'$ being an analog of $\gamma_L'$ for the right-handed quark $f$.}:
\begin{equation}
	\gamma_Q \to \gamma_Q' = \gamma_Q + \gamma_L', \qquad \beta_{Y_f} \to \beta_{Y_f'} = \beta_{Y_f} + \gamma_L'.
	\label{eq:gam_bet_shift}
\end{equation}
With the chosen $\mathcal{Z}_{Q_L}$ the RG functions \eqref{eq:gam_bet_shift} are affected already at the two-loop level.
The three-loop beta-functions can also be easily modified by adding the $h^3/\epsilon$ terms to $\mathcal{Z}_{Q_L}$.
However, having in mind the freedom \eqref{eq:equiv_classes}, it is easy to convince oneself 
	that it is possible to get rid of $\mathcal{Z}_{Q_L}$ together with arbitrary right-handed $\mathcal{Z}_{u}$, $\mathcal{Z}_{d}$ factors
	by the formal substitution $Y_f \to Y_f' = \mathcal{Z}^\dagger_{Q_L} Y_f \mathcal{Z}_{f}$  
	accompanied by the $SU(2)$-compatible change of the basis for the quark fields $Q^L \to Q^{'L} = \mathcal{Z}^\dagger_{Q_L} Q^{L}$, etc. 
As a consequence, the beta-functions $\beta_{Y_f}$ and $\beta_{Y_f'}$ from Eq.~\eqref{eq:gam_bet_shift} are equivalent, leading 
 	to the same RG flow of the quark sector ``observables'' --- six eigenvalues of $\YMu$ and $\YMd$ together
	with four independent parameters of the CKM matrix. 
Due to this, we restrict ourselves to the ``minimal'' case with all $\mathcal{Z} \equiv 1$, for which
	the quark anomalous dimensions are purely hermitian.
We would like to stress that this prescription is just a convenience choice. 
In order to justify this statement, one can use the analogy with the Hydrogen atom in a uniform magnetic field (see~Refs.~\cite{Grossman:2010gw,Grojean:2014pga}). 
Let us assume that the latter depends on some external parameter (an analog of scale $\mu$).  
By tuning the parameter one changes the field magnitude and its orientation with respect to some chosen basis. 
However, if one relates the components of the magnetic field to a measurable quantity (e.g., Zeeman splitting),  
	the dependence on the orientation drops out at any value of the considered parameter.
The same happens with the \MS~Yukawa matrices in the SM if we relate them (by matching procedure at certain scale $\mu$) 
	to some flavour (pseudo)observables.

To conclude, by explicit calculation we obtained the three-loop RGE for the general complex Yukawa matrices. 
The two-loop part reproduces the known expressions\footnote{One should identify $Y_u$, $Y_d$ and $Y_l$ 
	with $\mathbf{H^+}$, $\mathbf{F^+_d}$ and $\mathbf{F^+_L}$ of Refs.~\cite{Machacek:1983fi,Luo:2002ey}, respectively.} 
	\cite{Luo:2002ey}.  
The three-loop contributions are 
	free from gauge-parameter dependence and coincide with our previous results in the limit 
	of diagonal Yukawa couplings. 
In addition, we analyzed the ambiguity in the $\MS$ Yukawa matrix beta-functions, which appears starting from the two-loop level. 

In order to save space, 
we do not present the full expressions for $\beta_{Y_f}$, $f=u,d,l$ here.\footnote{The results in a computer-readable form are available as the ancillary files of the arXiv version of the paper.}
However, in a quite reasonable limit of vanishing couplings $g_1 = g_2 = Y_l = 0$ the beta-functions are not very lengthy, so
	we present here the result for $\beta_{Y_u} = \beta^{(1)}_{Y_u} + \beta^{(2)}_{Y_u} + \beta^{(3)}_{Y_u} + \dots$ in this approximation.
	Employing the notation \eqref{eq:yukawa_not} the loop expansion of $\beta_{Y_f}$ can be given as ($\lam = a_\lambda$)
{\allowdisplaybreaks
\begin{eqnarray}
	\beta_{Y_u}^{(1)} & = & 
-4 \aS
+\frac{3}{2}\left(\Yd+\Yu\right)
+\frac{3}{4}(\YMu-\YMd),
	\label{eq:yu_res_1}
	\\
	\beta_{Y_u}^{(2)} & = & 
3 \lam^2
-6 \lam \YMu
+\frac{11}{8}\YMdd
-\frac{1}{2}\YMdu
-\frac{1}{8}\YMud
+\frac{3}{4}\left(\YMuu+\Yud\right)
+\frac{15}{8}\YMd (\Yd+\Yu)
+8 \aS \left(\YMu-\YMd\right)
\nonumber\\
&&
+10 \aS \left(\Yd+\Yu\right)
-\frac{27}{8}\left(\Ydd+\Yuu + \left(\Yd + \Yu\right)\YMu  \right)
+\aS^2 \left(\frac{40}{9} \NGen-\frac{202}{3} \right),
	\label{eq:yu_res_2}
	\\
	\beta_{Y_u}^{(3)} & = & 
-18 \lam^3
- 
\aS^3
\left[1249 
-\left(
\frac{4432}{27}
-\frac{320}{3} \z3\right)
\NGen
-\frac{560}{81}  \NGen^2 
\right]
\nonumber\\
&&
+ 
\aS^2\left[
\Yd\left( \frac{457}{3} - 16 \NGen -108 \z3 \right) 
+\Yu\left(\frac{505}{3} - 16 \NGen -12 \z3 \right) 
\right.
\nonumber\\
&&\phantom{+\aS^2}~~ \left.
+\YMu \left(\frac{2779}{12} -11  \NGen -102 \z3 \right) 
-\YMd \left(\frac{2659}{12} -\frac{41}{3} \NGen -86 \z3 \right)
\right] 
\nonumber\\
&&
+\aS\left[ \YMdd \left( 13 + 40 \z3\right)
+  \YMud(14 + 4 \z3) 
- \YMdu (9 - 4 \z3)  
-38  \YMuu
+ 8 \lam \YMu 
+\Yud \left[\frac{57}{2} -24 \z3\right] 
\right.
\nonumber\\
&&\left.
+(\Yd+\Yu)
\left[ \YMd \left(\frac{97}{4} - 36 \z3\right)  - \left(\frac{177}{4} - 36 \z3\right)\YMu \right] 
+\left[\frac{15}{4} - 36 \z3\right] (\Ydd+\Yuu)  
\right] 
\nonumber\\
& & 
+\lam \left(\frac{3}{2} \YMud
-15 \YMdd
+\frac{63}{2}  \YMuu
+45 \YMu (\Yd+\Yu)
+\frac{45}{2}  (\Ydd+\Yuu)
\right) 
\nonumber\\
&&
+\lam^2 \left(
\frac{285}{8}  \YMu
-\frac{21}{8}  \YMd
-\frac{135}{4} (\Yd+\Yu)
\right)
+\left(\frac{9}{8} - \frac{9}{4} \z3 \right) \YMddd
-\left(\frac{345}{32} -\frac{9}{4} \z3\right) \YMuuu
\nonumber\\
&&
+\frac{43}{16} \YMudu
+\frac{75}{32} \YMuud
+\frac{83}{32} \YMduu
-\frac{37}{16} \YMdud
-\left(\frac{95}{16} - 6 \z3\right) \YMudd
-\left(\frac{183}{32} - 6 \z3\right) \YMddu
\nonumber\\
&&
+\left(\frac{3}{8} \YMdu 
-\frac{69}{8} \YMdd 
-\frac{9}{8} \YMuu 
+\frac{21}{8}\YMud\right) (\Yd+\Yu)
\nonumber\\
&&
+\left(\frac{135}{16} \YMu  - \frac{9}{4} \YMd \right) (\Yuu+\Ydd)
+\left(\left(\frac{147}{4} - 36 \z3 \right)\YMd 
-\frac{81}{8} \YMu\right) \Yud
\nonumber\\
&&
\nonumber\\
&&
+\left(\frac{789}{32} + \frac{9}{2} \z3 \right) (\Yddd+\Yuuu)
+\frac{831}{32} (\Yudd+\Yuud)
	\label{eq:yu_res_3}
\end{eqnarray}
}
	where $\NGen=3$ is the number of fermion generations. 
It is worth pointing out that the corresponding expressions for $\beta^{(l)}_{Y^\dagger_u}$ can be deduced
	from \eqref{eq:yu_res_1}, \eqref{eq:yu_res_2} and \eqref{eq:yu_res_3} 
	by the substitutions
\begin{equation}
	\YMud \leftrightarrow \YMdu, \quad 
	\YMudd \leftrightarrow \YMddu, \quad 
	\YMuud \leftrightarrow \YMduu.
	\label{eq:betaY_to_betaYc}
\end{equation}

The obtained expressions can be applied to RGE studies of different BSM models aimed to unveil the physics
	behind the observed SM flavour pattern.
It is also worth mentioning that from $Y_u$ and $Y_d$ it is possible 
to deduce the three-loop RGE for the CKM matrix elements \cite{Babu:1987im,Naculich:1993ah,Balzereit:1998id} 
or Quark Flavour invariants~(see Refs.~\cite{Jenkins:2009dy,Schwertfeger2014}) in the \MS-scheme.

\subsection*{Acknowledgments}
The authors would like to thank M.~Kalmykov, A.~Pivovarov, and S.~Schwertfeger for stimulating discussions.
This work is supported by the RFBR grant 12-02-00412-a.
The research of V.N.~Velizhanin is supported by a Marie Curie International Incoming Fellowship within the 7th European Community Framework Programme, grant number PIIF-GA-2012-331484 and by SFB 647 ``\emph{Raum -- Zeit -- Materie. Analytische und Geometrische Strukturen}''.
Additional support from Russian President Grant No. MK-1001.2014.2 and Dynasty Foundation is kindly acknowledged by A.V.~Bednyakov and A.F.~Pikelner.

\end{document}